\documentclass[
    ,final            
  ]
  {aipproc}

\layoutstyle{6x9}
\newcommand{\be}{\begin{equation}}
\newcommand{\ee}{\end{equation}}
\newcommand{\bea}{\begin{eqnarray}}
\newcommand{\eea}{\end{eqnarray}}

\newcommand{\bc}{\begin{center}}
\newcommand{\ec}{\end{center}} 
\newcommand{\g}{\gamma} 
\newcommand{\f}{\frac} 
\newcommand{\bi}{\begin{itemize}}

\newcommand{\n}{\nonumber \\} 
\def\ba{\begin{eqnarray*}}
\def\ea{\end{eqnarray*}}
 
\begin{document}

\title{Traveling waves in high energy QCD}


\author{Robi Peschanski}{
  address={Service de physique th{\'e}orique, CEA/Saclay,
  91191 Gif-sur-Yvette cedex, France\\
  URA 2306, unit{\'e} de recherche associ{\'e}e au CNRS}}

\begin{abstract}
Saturation is  expected to occur when a high density of partons (mainly 
gluons)- or equivalently strong fields in Quantum Chromodynamics (QCD) - is 
realized in the weak coupling regime. A way to reach saturation  is through the 
high-energy evolution of an extended target probed at a fixed hard scale. In 
this case, the transition to saturation is expected to occur from nonlinear 
perturbative QCD  dynamics. We discuss this approach to saturation, which is 
mathematically characterized by the appearance of traveling wave patterns in a 
suitable kinematical representation. A short review on traveling waves in high 
energy QCD and a first evidence of this phenomenon in deep-inelastic proton 
scattering are presented.
\end{abstract}

\maketitle


\section{Introduction}

One of the most intringuing  empirical observations about deep-inelastic 
scattering (DIS) on a proton is the geometric scaling property 
\cite{Stasto:2000er}, 
which states that the ${\g ^*\!-\!p}$  total cross-section data
can be approximately ploted on a one-dimensional curve  
$\sigma^{\g *\!-\!p}(Q^2/Q_s^2(Y))$ where $(Q^2,Y)$ span the 
virtuality-rapidity 
 kinematical domain of DIS. One finds
$Q_s^2(Y) \approx e^{\lambda Y}.$ 

The question we want to address is whether the observed geometric scaling can 
be 
explained in terms of  QCD evolution equations describing {\it saturation}.

Saturation of parton (mainly gluon) densities
at high rapidity may be at the origin of the geometric scaling 
property. Saturation is expected to occur when a high density of 
partons is created in the limited geometry of the target. In the 
expected scenario, the parton  wave functions overlap and eventually create a 
new state of matter,  the Color Glass Condensate (CGC)  \cite{Itakura:2005eu}. 
In the case of DIS on a proton, increasing densities are obtained by 
high-energy evolution at fixed virtuality $Q^2,$ when the corresponding linear 
QCD evolution, named BFKL after Balitsky, Kuraev, Lipatov, and Fadin 
\cite{BFKL}, gives rise to an exponential growth of the number of gluons. At 
high enough energy the  BFKL formulation gets modified. In the   approximation 
of uncorrelated hard probes, one deals with the 
Balitsky-Kovchegov (BK) equation \cite{bk}, where  the corrections due to the 
high density of partons are expressed in terms of a non-linear damping term. We 
shall investigate whether and how geometric scaling could be due to the 
nonlinear  QCD evolution of the gluon distribution in the target.

\section{Geometric scaling as traveling waves}

\begin{figure}[bt]
\includegraphics[width=8cm,angle=-90]{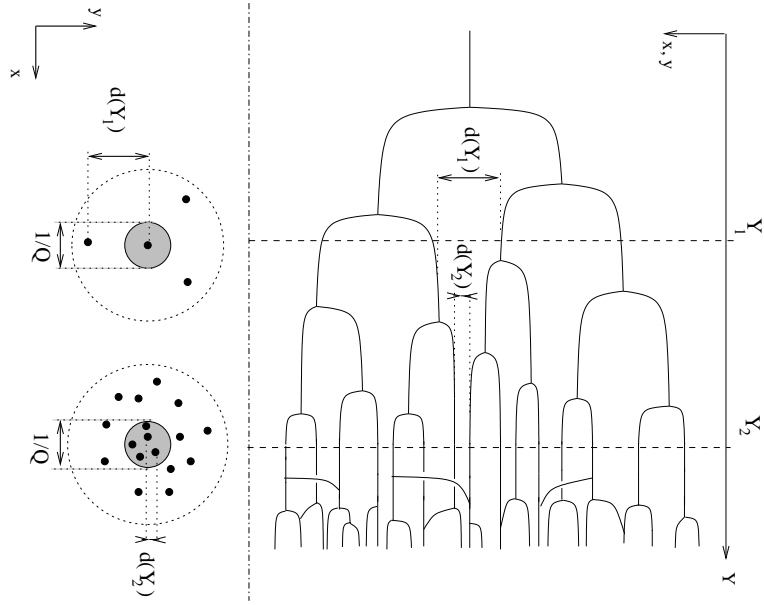}
\caption{Schematic picture of the  transition to saturation with rapidity. In a 
region $Y \sim Y_1:$ {the projectile  of size $\sim 1/Q$  is able to probes the 
 number of partons in  exponential growth}; when $Y \sim Y_2:$ the probe  
counts  partons by groups, leading to a nonlinear damping evolution factor in 
the probed parton density. For $Y > Y_2:$ A new phase of partonic matter (CGC) 
may appear through parton correlations.}
\label{eikonal}
\end{figure}

Let us start from the BK equation expressed in momentum space (using the  
impact-parameter independent formulation)
$${{
{\partial_Y}{\cal N}=\bar\alpha
\chi\left(-\partial_L\right){\cal N}
-\bar\alpha\, {\cal N}^2 
}}$$
where ${\cal N}(Y,L=2\log k)$ is related to the tranverse-momentum $k$ 
distribution of gluons in the target. The (leading order) BFKL kernel is
$$
\chi\left(-\partial_L\right)=2\psi(1)
-\psi\left(-\partial_L\right)-\psi\left
(1\!+\!\partial_L\right)\ .
$$
The coupling constant will  be considered either  constant, or running like 
$\bar\alpha = 
1/bL.$ 

From a mathematical point of view, the BK equation can be related \cite{tv} to 
an ``universality class'' of  nonlinear equations for which characteristic 
properties can be rigorously derived. ``Universality'' means here that the 
solutions have general characteristics which are essentially independent from 
initial conditions. In the case of the BK equation (and also for some 
extensions beyond BK) this generic feature is the formation of {\it traveling 
wave} patterns when the energy increases. As we shall see, traveling waves are 
intimately related to geometric scaling. Let us define these traveling waves 
and give a qualitative explanation of their formation during the rapidity 
evolution. 

Consider first the ``diffusive approximation'' of the BFKL 
kernel$$
\bar\chi\left(-\partial_L\right)
\sim \chi\left({\scriptstyle \f 12}\right)+
{\scriptstyle \f 12}\ {\scriptstyle  \chi''\left({\scriptstyle \f 12}\right) 
}\times 
\left(\partial_L+{\scriptstyle \f 12}\right)^2\ .$$ By a suitable redefinition 
of the 
function and variables one can map the BK equation in the  diffusive 
approximation onto 
the Fisher Kolmogorov-Petrovski-Piskounov (F-KPP) equation \cite{KPP} 
describing a 
reaction-diffusion process in space and time: 
$${{
\partial_t u(t,x)=\partial_x^2 u(t,x)+u(t,x)-u^2(t,x)}}\ , $$
In reaction-diffusion language, $\partial_x^2 u(t,x)$ is the diffusion term, 
$u(t,x)$ is responsible for the  exponential growth and $u^2(t,x),$ the damping 
term.

The ``dictionnary'' between F-KPP and BK can be written as follows
\ba
{Time} \ {t}\ &\to &   Y\ \ \n 
{Space} \ {x}\ &\to & 
L+{\scriptstyle \f 12}\ {\bar\alpha \chi''\left({\scriptstyle \f 12}\right)}\ 
Y\ \ \n 
{Wave\ Front} \ {\ u(x-ct)} &\to & \    {\cal N}(L-vY)\ \n 
{Traveling\  Waves} \ &\to & {Geometric\  Scaling}.
\ea 

\begin{figure}[t]
\includegraphics[width=10cm]{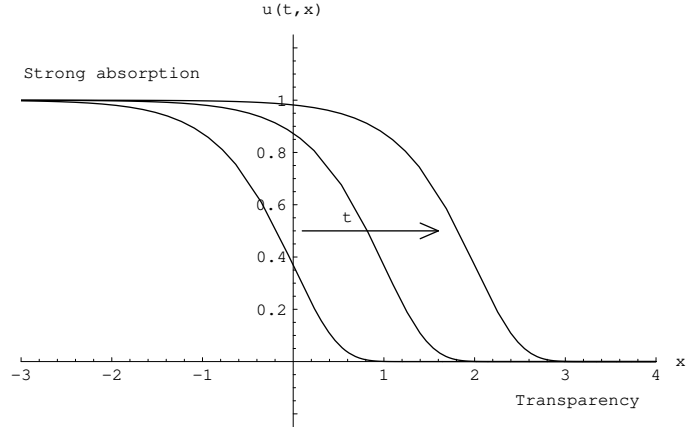}
\caption{Typical traveling wave pattern. The wave front (in QCD: amplitude) 
connecting the regions $u=1$ (in QCD: strong absorption) and $u=0$ (in QCD: 
transparency) travels from the left to the right as $t$ (in QCD: $Y$) 
increases.}
\label{dis}
\end{figure}

The key feature of the solutions of the F-KPP equation (and of equations 
belonging to the same universality class) is the formation of traveling wave 
patterns, see Fig.(\ref{dis}). It can be rigorously proven that, at large times 
($\Rightarrow$ large rapidities), the solutions verify  
$$u(t,x)\f
{}{t\rightarrow\!\infty}\!\!>u(x\!-\!m(t)) 
\Rightarrow
{{{\cal N}(Y,L)\ \f
{}{Y\rightarrow\!\infty}\!\!>\ {\cal N}\left(L\!-\!{L_s(Y)}\right)}}$$
with  $${{L_s(Y)=\log 
Q_s^2(Y) =
\bar\alpha \  {\chi'(\bar\gamma)} Y-\f {3}{2\bar\gamma}\log Y
-\f {3}{(\bar\gamma)^2}
\sqrt{\f {2\pi}{\bar\alpha\chi^{\prime\prime}(\bar\gamma)}}\f {1}{\sqrt{Y}}
\ +\ {\cal O}(1/Y)
}}\ . $$ 
$\bar\gamma$ is the implicit solution of the characteristic equation 
${\chi(\bar\gamma)}/{\bar\gamma}=\chi'(\bar\gamma)$ which  leads to the 
asymptotic wave speed $v=\bar\alpha \chi'(\bar\gamma).$ It plays the r\^ole of 
a critical QCD 
anomalous dimension. One immediately recognizes the equivalence between 
geometric scaling and traveling waves  solutions of BK. Moreover, quite a 
detailed information can be obtained from the sole knowledge of the  linear 
kernel, as for  the three first terms \cite{tv} of the asymptotic expansion of 
the saturation scale $Q_s^2(Y).$

\section{Asymptotic traveling waves}

In fact the results using the diffusive approximation of BK and its mapping to 
the F-KPP equation are more general. Instead of entering here in refined (and 
technically useful) mathematical arguments \cite{brunet}, let us qualitatively 
illustrate the reason why traveling wave patterns generically develop during 
the rapidity evolution from a given initial condition, using the concepts of 
``pulled fronts''.
\begin{figure}[t]
\includegraphics[width=8cm]{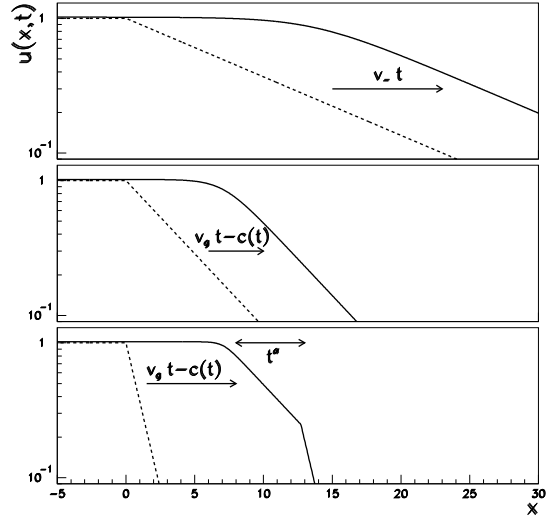}\caption{``Pulled vs. Pushed fronts''.
The function $u(t,x)$ is
represented for the three different classes of initial conditions. Top: 
``Pushed fronts,'' 
the wave front keeps the initial front profile with a non-universal speed 
$v_-;$ Middle:   ``Pulled = Pushed''; Bottom: ``Pulled  front:'' both the wave 
front and the speed acquire universal values at large $t\ (\sim Y)$ in an 
intermediate kinematical region called the ``wave interior''.}
\label{front}
\end{figure}

Universality features are obtained in the ``pulled'' front situation, see the 
bottom plot of 
Fig.(\ref{front}). In this case, one starts with initial conditions 
corresponding to an anomalous dimension $\g_0 > \bar \g,$ where $\bar 
\g=.6275... $ is  for the (leading order) BFKL kernel \cite{tv}.
\begin{figure}[t]
\includegraphics[width=11cm]{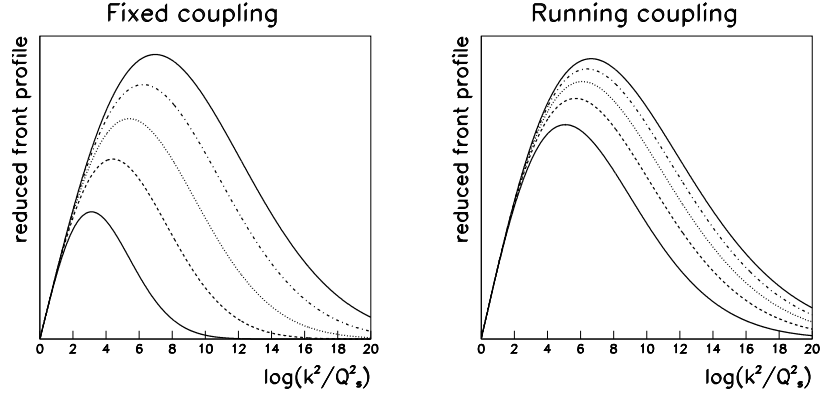}
\caption{The ``Reduced'' Front profile.}\label{reduced}
\end{figure} The formation of the traveling wave comes from the competition 
between the 
exponential growth at small values of the solution $u,$ where the linear term 
dominates 
the evolution, with the   nonlinear damping exercised at larger values of $u.$ 
Constrained by these two opposite trends, the solution is  ``forced'' to adopt 
an universal behaviour. Fortunately enough, the perturbative QCD initial 
conditions, through the transparency property ($\g_0 =1 > \bar 
\g,$), fall into the universal ``pulled'' front regime \cite{tv}.
 
In Fig.(\ref{reduced}) the ``reduced front'' 
$(k^2/Q_s(Y)^2)^{\bar\gamma}\,{\cal N}(L,Y),$ obtained  by factorizing out the 
wave propagation, is 
represented for fixed (left) and running (right) coupling constant. It shows 
the scaling straight line, which  envelopes  the curves obtained for increasing 
rapidity. Hence geometric scaling violations are also predicted  due to 
diffusion. One may notice that the diffusive approach to scaling is slower for  
running coupling, corresponding to ``anomalous diffusion'' in $1/t^{1/3}\sim 
1/Y^{1/6}$ (for running coupling  $t \sim Y^{1/2}$). Normal diffusion  
$1/t^{1/2}\sim 1/Y^{1/2}$ is characteristic of the F-KPP universality class or 
BK with fixed coupling.

\section{Parametric traveling waves}

The abovementionned mathematical properties have the major interest of 
revealing the link between geomeric scaling and the nonlinear dynamics of QCD 
evolution equations at high energy. However the path from phenomenology to 
theory is not yet accomplished. On the phenomenological ground, the simulations 
of BK equation solutions \cite{Enberg:2005cb} have shown that pre-asymptotics 
effects may be important and endenger   geometrical scaling predictions from 
QCD in the physical region where it is observed. On 
the theory side, it has been recently realized \cite{Mueller:2004se} that  the 
effect of parton fluctuations and correlations, especially in the 
dilute (transparency) region which is the driving force of the pulled front,  
may be responsible for a  breaking of geometric scaling. In fact, traveling 
wave patterns still exist \cite{Iancu:2004es} but their event-by-event  
fluctuations tends to break geometric scaling in the average.

Despite these problems, the numerical simulations \cite{Enberg:2005cb} also 
show that  patterns behaving like traveling waves  exist in a preasymptotic  
region in energy. Moreover the mean field BK equation seems to be still valid 
in this region. In order to explore this possibility and understand its origin, 
we proposed \cite{parametric} to attack the problem of traveling wave solutions 
of the BK equation in a new way.

The initial mathematical idea \cite{logan} is to assume and  directly
 insert a traveling wave solution {$u(x,t)\to U(s= x/c -t)$} into the F-KPP 
equation, or equivalently  to impose  a geometric scaling form to a solution of 
the BK equation. Keeping for simplicity the space-time language and the FKPP 
equation (the dictionnary to BK being still easy), one gets
$$
{{U(1-U)+\f {dU}{ds}+ \f1{c^2} \f {d^2U}{(ds)^2} =0}}\ .
$$ 
taking into account the  previous study showing \cite{KPP} that  the wave speed 
$c\ge 2,$  we obtain an iterative solution 
$$U(s) = U_0+ {\f 1{c^2}}U_1+\sum_{p\ge2} {\f 1{c^{2p}}}U_{2p} 
$$
obeying an exactly solvable \cite{parametric} hierarchy of equations for the  
$U_{2p}$'s. 
The first one  only (for $U_0$) is nonlinear and can be  exactly solved, the 
other boiling down to a rather simple linear algebra.
This mathematical result translates into a {\it parametric}  form of a 
geometric scale invariant gluon amplitude ${\cal N}(L-L_s(Y))$ which can be 
used in phenomenology.  ${\cal N}$ depends only on few parameters determined by 
the linear kernel. One obtains $$
{\cal N}\propto \f 1{1\!+\!\left[\f {k^2}{Q^2_s(Y)}\right]^{\mu}}\!-\!\f 
1{c^2}\ \f 
{\left[\f{k^2}{Q^2_s(Y)}\right]^{\mu}}
{\left(1\!+\!\left[\f{k^2}{Q^2_s(Y)}\right]
^{\mu}\right)^2}\  
\log \f {\left(1\!+\!\left[\f {k^2}{Q^2_s(Y)}\right]^{\mu}\right)^2}
{4\left[\f{k^2}{Q^2_s(Y)}\right]^{\mu}}\ . $$
 
This solution  (valid also for the running case with an appropriate mapping 
\cite{parametric}) is mathematically valid \cite{brunet} in  the ``wave 
interior'', see Fig.(\ref{front}). Remarkably, we observed that the 
corresponding region 
has a large overlap with the physical region where geometrical scaling is 
valid at HERA. In order to check this opportunity, we determined  the 
phenomenological 
observable related to ${\cal N} (L,Y)$ from a MRST parametrization of $F_2$ 
data.
 \begin{figure}[t]
\includegraphics[width=11cm]{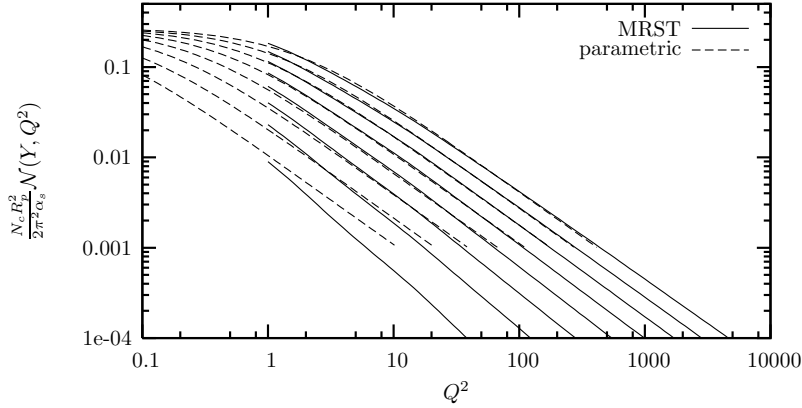}
\caption{The amplitude ${\cal N}(L,Y)$ as a fonction of $L$ for different 
values 
of $Y\!=\!3,4,..,10.$ Full lines:   MRST gluon parametrization. Dashed lines: 
Parametric QCD traveling waves.} 
\label{fig4}
\end{figure}
The obtained kernel parameters are not far but different from those expected 
from the leading order BFKL kernel one and thus could require higher order 
contributions \cite{parametric}. Interestingly, higher order contributions 
plays an important r\^ole in a different but related problem: the existence of 
``backward'' traveling waves appearing \cite{enberg}  from the energy evolution 
starting from  an inital  high density (and not initial low density as in DIS) 
state, which could be the case in  heavy-ion reactions.

 \section{Conclusion}
Traveling wave patterns have been shown to result from nonlinear QCD equations. 
The 
phenomenologically observed geometric scaling in DIS on a proton 
at large energy is consistent with the traveling wave patterns. It remains to 
be found 
whether one can relate the geometric scaling curve with the QCD kernel and more 
generally to a  complete solution of high-energy QCD, including correlation and 
fluctuation contributions.


\begin{theacknowledgments}
I wish to take the opportunity of the  kind hospitality from  the organizers of 
ISMD 2005 
to acknowledge the long-term interactions with   Andrzej Bialas, Jean-Louis 
Meunier  and  
Bernard Derrida, which  inspired me to  look for properties of nonlinear 
equations such as F-KPP. I also acknowledge the  fruitful collaboration with 
St\'ephane Munier during  the 
first stage of the present study and, more recently, a stimulating cooperation 
with 
Rikard Enberg, Cyrille Marquet and Gregory Soyez. Lively discussions with 
Edmond Iancu, 
Al Mueller and many other colleagues have to be mentionned. 
\end{theacknowledgments}



\end{document}